\documentclass[10pt,conference,letterpaper]{IEEEtran}
\usepackage{amsmath,amsfonts}
\usepackage{algorithmic}
\usepackage{algorithm}
\usepackage{array}
\usepackage{textcomp}
\usepackage{stfloats}
\usepackage{url}
\usepackage{verbatim}
\usepackage{graphicx}
\usepackage{cite}
\usepackage{multicol}
\usepackage{xspace}
\usepackage{xcolor}
\usepackage{hyperref}
\usepackage{colortbl}
\usepackage[bf]{caption}
\usepackage{subcaption}
\usepackage{enumitem}
\usepackage{soul}
\usepackage{multirow}
\usepackage{todonotes}
\usepackage{balance}

\newcommand{\proj}{\emph{DataStates-LLM}\xspace}

\begin{document}
\title{\proj: Scalable Checkpointing for Transformer Models Using Composable State Providers}

\author{
\IEEEauthorblockN{Avinash Maurya\IEEEauthorrefmark{1},
M.~Mustafa Rafique\IEEEauthorrefmark{2},
Franck Cappello\IEEEauthorrefmark{1},
Bogdan Nicolae\IEEEauthorrefmark{1}}
\IEEEauthorblockA{\IEEEauthorrefmark{1}Argonne National Laboratory, Lemont, IL, USA; \texttt{\{amaurya,cappello,bnicolae\}@anl.gov}}
\IEEEauthorblockA{\IEEEauthorrefmark{2}Rochester Institute of Technology, Rochester, NY, USA \texttt{mrafique@cs.rit.edu}}
}

\maketitle

\begin{abstract}

The rapid growth of Large Transformer-based models, specifically Large Language Models (LLMs), now scaling to trillions of parameters, has necessitated training across thousands of GPUs using complex hybrid parallelism strategies (e.g., data, tensor, and pipeline parallelism). Checkpointing this massive, distributed state is critical for a wide range of use cases, such as resilience, suspend-resume, investigating undesirable training trajectories, and explaining model evolution. However, existing checkpointing solutions typically treat model state as opaque binary blobs, ignoring the ``3D heterogeneity'' of the underlying data structures--varying by memory location (GPU vs. Host), number of ``logical'' objects sharded and split across multiple files, data types (tensors vs. Python objects), and their serialization requirements. This results in significant runtime overheads due to blocking device-to-host transfers, data-oblivious serialization, and storage I/O contention. 

In this paper, we introduce \proj, a novel checkpointing architecture that leverages \textit{State Providers} to decouple state abstraction from data movement. \proj exploits the immutability of model parameters during the forward and backward passes to perform ``lazy'', non-blocking asynchronous snapshots. By introducing State Providers, we efficiently coalesce fragmented, heterogeneous shards and overlap the serialization of metadata with bulk tensor I/O. We evaluate \proj on models up to 70B parameters on 256 A100-40GB GPUs. Our results demonstrate that \proj achieves up to 4$\times$ higher checkpointing throughput and reduces end-to-end training time by up to 2.2$\times$ compared to state-of-the-art solutions, effectively mitigating the serialization and heterogeneity bottlenecks in extreme-scale LLM training.
\end{abstract}


\section{Introduction}
\label{sec:intro}
Large Language Models (LLMs) have become a cornerstone of modern artificial intelligence (AI), driving unprecedented capabilities in text generation, comprehension, and domain-specific scientific discovery~\cite{gottweis2025aicoscientist}. To achieve these capabilities, model sizes have exploded, routinely exceeding hundreds of billions to trillions of parameters~\cite{llama3training}. Training such massive models requires High-Performance Computing (HPC) infrastructure comprising thousands of GPUs and involves running for weeks or months~\cite{workshopBLOOM176BParameterOpenAccess2023,gottweis2025aicoscientist}. To manage the memory footprint of these models, training runtimes employ complex parallelism strategies: they combine Data Parallelism (DP), Tensor Parallelism (TP), and Pipeline Parallelism (PP), along with optimizer state sharding techniques like ZeRO~\cite{rasleyDeepSpeedSystemOptimizations2020} and FSDP~\cite{zhao2023pytorch} .
(e.g., BLOOM~\cite{workshopBLOOM176BParameterOpenAccess2023} and Llama~3~\cite{llama3training}).

\paragraph*{\textbf{Motivation: The Need for Scalable Checkpointing}} 
Given the large scale and extended duration of LLM training, checkpointing is a fundamental primitive necessary to ensure resilience and productivity. Hardware failures, software bugs, and timeouts are statistically inevitable at scale~\cite{robust-bytedance}. Without frequent checkpointing, the amount
of lost computations that need to be recomputed is too costly both in terms of time and resources. For example, the Llama~3 405B model training involved 16K GPUs for 54 days and encountered failures every 2.8 hours~\cite{llama3training,robust-bytedance}. Another example is Alibaba's Unicron training that has a reported failure rate of 43.4\%~\cite{he2023unicron}.

Beyond resilience, checkpoints are essential for addressing training instabilities (e.g., loss spikes in PaLM and GLM-130B~\cite{takase2023spike} models), which are hard to predict and defend against, leaving rollback followed by adjustments as the most viable correction strategy. The productivity of several other prominent scenarios also depends on scalable checkpointing: reinforcement learning from human feedback~(RLHF), transfer learning,
faster convergence by merging different model states along the training trajectory.
In such cases, the checkpoint frequency can be as high as every iteration.

Regardless of the scenario, checkpointing involves the fundamental ability to capture a distributed AI model state frequently at scale without incurring significant overheads (i.e., blocking) that interrupt applications from making progress.

\paragraph*{\textbf{Challenges and Limitations of State-of-the-Art}}
Compared to conventional deep-learning models (ResNet, VGG, etc.), which range in hundreds of MBs and fit on a single GPU memory, LLMs are routinely composed of billions of parameters, leading to \emph{massive checkpoint volumes}. The large model and optimizer states need to consistently combine multiple data structures of \emph{different data types and sizes}, such as tensors, arrays, and custom Python objects (e.g., dictionaries, seeds of PRNGs, etc.). These
data structures may be generated across different runtimes and languages,
and may live in different memory tiers (e.g., Python dictionaries held in host memory, C++/CUDA tensors held in GPU memory, etc.). Furthermore, the combination of data-, pipeline-, and tensor-parallelism, along with redundancy elimination approaches (e.g., ZeRO~\cite{rasleyDeepSpeedSystemOptimizations2020}, FSDP~\cite{zhao2023pytorch}) results in a fine-grain distribution of
these data structures across a large number of compute nodes. We call these emerging multi-faceted variations in data types, data sizes, and different data sharding/distribution strategies across different storage tiers and compute
nodes \emph{3D heterogeneity of LLM checkpoints}. This aspect is insufficiently addressed by state of art checkpointing approaches~\cite{wang2024fastpersist, maurya2024datastates, TorchSnapshot, mohanCheckFreqFrequentFineGrained, wang2023gemini}, which results in significant overhead during checkpointing.

Specifically, many checkpointing approaches implemented in state-of-the-art LLM training runtimes (e.g., DeepSpeed~\cite{rasleyDeepSpeedSystemOptimizations2020}) deal with data sharding by initiating checkpoint capture in parallel from all GPUs to saturate I/O bandwidth. However, they do so in a blocking fashion, which interrupts the critical training path. Alternatives such as multi-level asynchronous checkpointing techniques aim to mask these overheads by capturing checkpoints on a fast tier, then flush the checkpoints from the fast tier to slower tiers in background~\cite{nicolaeDeepFreezeScalableAsynchronous2020,mohanCheckFreqFrequentFineGrained,wang2024fastpersist}. However, it is not straightforward to adopt these techniques directly in LLM runtimes because of several reasons. First, GPUs don't typically have enough spare memory capacity to capture full  (CheckFreq~\cite{mohanCheckFreqFrequentFineGrained}, GEMIMI~\cite{wang2023gemini}). Second, although it is possible to capture the checkpoints directly on the host memory (e.g., TorchSnapshot~\cite{TorchSnapshot}, CheckFreq~\cite{mohanCheckFreqFrequentFineGrained}), the limited GPU to host PCIe transfer bandwidth is orders of magnitude slower than GPU memory bandwidth
and also needs to be shared by multiple GPUs typically co-located on the
same compute node, thus incurring high I/O overheads. Without better techniques
to capture a checkpoint on the initial fast tier, the benefit of multi-level
asynchronous checkpointing is greatly reduced, to the point where it is not
significantly faster than synchronous checkpointing. For example, despite the availability of high speed links (50+~GB/s network and 25+~GB/s PCIe), the LLM checkpointing throughput is far from saturating the link capacity (e.g., REFT~\cite{wang2023reliableREFT} reports 38\% saturation), and often drops as low as a few GB/s (e.g., as reported by Nebula~\cite{ziqiwangOptimizeCheckpointPerformance2023}). 

Heterogeneity of data generated by different runtimes and programming languages on different memory tiers is also a problem. The majority of checkpointing approaches
mentioned above are highly optimized for tensor capture and serialization,
but neglect other data structures. Therefore, training runtimes, such as DeepSpeed~\cite{rasleyDeepSpeedSystemOptimizations2020}, typically collect and capture high-level metadata and other Python data structures separately in a centralized fashion, before capturing the content of the distributed tensors in parallel. At scale, this blocking step becomes a significant bottleneck, even if the size of high-level metadata is 
significantly less than 
the size of the tensors. 

\paragraph*{\textbf{Key Insights and Contributions}}
In this paper, we propose \proj, a high-performance asynchronous checkpointing system specifically designed for \emph{3D checkpoint heterogeneity}. Our approach is built on two key insights. First, model parameters and optimizer states remain \textit{immutable} during the compute-heavy forward and backward passes of a training iteration. This creates a window of opportunity to perform ``lazy'' Device-to-Host (D2H) copies without using intermediate staging areas or blocking I/O transfers and computations, which solves the limitation of multi-level asynchronous techniques faced by state of art LLM checkpointing approaches.
Second, allowing the checkpointing runtime to be aware of all data structures
regardless of their type or where they live opens an opportunity to parallelize and reorder operations better (e.g., flush the host-resident data structures to disk while the GPU-resident ones are flushed to the host).

This paper capitalizes on the two key insights and extends our previous work~\cite{maurya2024datastates} 
to design and implement a checkpointing runtime specifically optimized to efficiently handle 3D checkpoint heterogeneity at scale. By introducing \emph{composable state providers}, which are complementary abstractions optimized for capturing specific aspects of heterogeneity, we enable efficient asynchronous streaming into a globally consistent state that represents a full checkpoint. We summarize our contributions as follows:

\begin{enumerate}[leftmargin=*]
    \item \emph{Gap Analysis of LLM Checkpointing:} We quantify the impact of 3D parallelism on checkpoint composition, highlighting the ``3D checkpoint heterogeneity'' of data structures (GPU vs. Host, Tensors vs. Objects) and identifying serialization as a critical blocking bottleneck (\S~\ref{sec:motivation}).
    
    \item \emph{Design of State Providers:} We introduce \textit{State Providers}, a middleware abstraction that encapsulates the semantics of heterogeneous data structures. This allows \proj to perform zero-copy serialization for tensors while efficiently handling complex Python objects, solving the state heterogeneity challenge (\S~\ref{sec:design}).
    
    \item \emph{Lazy, Non-Blocking Asynchrony:} We implement a lazy state capture mechanism that overlaps D2H transfers with the immutable phases of training (forward/backward passes), which hides the cost of state capture (\S~\ref{sec:design:principles:lazy}).
    
    \item \emph{Streamlined Multi-Tier Kernel-Accelerated I/O Engine:} We design a pipelined I/O engine that manages pinned host memory pools and performs kernel-accelerated, multi-threaded flushing to persistent storage using low-latency I/O libraries, such as \texttt{liburing}, thus maximizing I/O bandwidth utilization (\S~\ref{sec:design:principles:streamlined}).
    
    \item \emph{Scalability Evaluation and I/O Overlap Profiling:} We evaluate \proj on the Polaris supercomputer, training Llama~2 models up to 70B parameters on 256 A100-40GB GPUs, and performing ablation studies to investigate the gains of each design proposal. We demonstrate a 3$\times$--4.2$\times$ improvement in checkpointing throughput and a 1.3$\times$--2.2$\times$ reduction in end-to-end training time compared with TorchSnapshot, a state-of-the-art approach, and with our own previous work (\S~\ref{sec:evaluation}).
\end{enumerate}

\section{Background}
\label{sec:background}

\paragraph*{\textbf{3D Parallelism: Data, Pipeline, and Tensor Strategies}}
Data parallelism (DP) remains the foundational technique for accelerating deep learning by replicating models across multiple workers working on independent minibatches~\cite{rajbhandari2021zero, workshopBLOOM176BParameterOpenAccess2023, llama3training}. To maintain replica consistency, gradients are synchronized via all-reduce collectives before the model update phase. However, as LLM footprints exceed single-GPU capacity, hybrid 3D parallelism is required. Pipeline parallelism (PP) partitions the model vertically into stages composed of sequential transformer layers. To maximize throughput and GPU occupancy, minibatches are divided into microbatches, allowing forward and backward passes to overlap across stages in a pipelined fashion. Tensor parallelism (TP) provides horizontal sharding by distributing individual transformer blocks and associated memory across multiple GPUs~\cite{shoeybiMegatronLMTrainingMultiBillion2020}. Due to high intra-layer communication overheads, TP is typically confined to node-local GPUs interconnected via high-speed fabrics. At extreme scales, these three strategies are combined into \emph{3D parallelism} to balance memory efficiency and communication latency.

\begin{figure*}[t]
    \centering
    \includegraphics[width=0.99\linewidth]{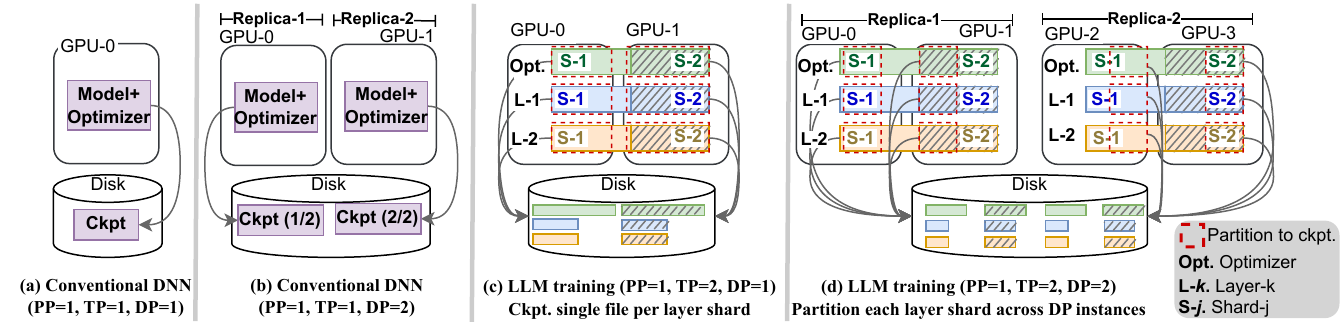}
    \caption{Sharding of checkpoints during AI model training for pipeline (PP), tensor (TP), and data (DP) parallelism.}
    \label{fig:ckpt-sharding}
\end{figure*}

\paragraph*{\textbf{Redundancy Elimination via State Sharding}}
The redundancy inherent in DP replicas can be mitigated by sharding model states across workers, such that each worker is responsible for the management of a unique shard. Runtimes like DeepSpeed~\cite{rajbhandariZeROMemoryOptimizations2020} and FSDP~\cite{zhao2023pytorch} implement optimization stages (e.g., ZeRO Stage-1/2/3) to shard the optimizer state, gradients, and model parameters, respectively. While sharding improves memory efficiency by reducing the aggregated amount of GPU memory required to accommodate the model, it requires collective communication to reconstruct full states during forward/backward passes and complicates state management during checkpointing.

\paragraph*{\textbf{Implications of State Sharding on Checkpointing}}
In the case of training on a single GPU, the model and optimizer states are serialized into a monolithic file (Figure~\ref{fig:ckpt-sharding}(a)). DP allows for I/O parallelization by assigning different shards to be checkpointed by independent workers (Figure~\ref{fig:ckpt-sharding}(b)), an approach utilized by systems like DeepFreeze~\cite{nicolaeDeepFreezeScalableAsynchronous2020} and TorchSnapshot~\cite{TorchSnapshot}. For LLMs, sharding extends beyond DP to individual model layers, enabling parallel writes even without data parallelism (Figure~\ref{fig:ckpt-sharding}(c)). The integration of 3D parallelism further partitions these layer shards across DP instances (Figure~\ref{fig:ckpt-sharding}(d)). Modern runtimes, e.g., DeepSpeed, leverage this high degree of sharding to maximize I/O throughput on parallel file systems; however, this results in an explosion of independent files, leading to metadata server bottlenecks~\cite{Gossman2026LLMCheckpointIO}. This work adopts the default DeepSpeed strategy of sharding into separate files, focusing on accelerating the capture and transfer of these distributed objects.

\section{Related Work}
\label{sec:related}

\paragraph*{\textbf{Checkpointing in Deep Learning}}
Systems such as CheckFreq~\cite{mohanCheckFreqFrequentFineGrained} aim at performing fine-grained iteration-level checkpoints and overlap checkpoint flushes with the training phases, but do not support checkpointing in pipeline parallel training setups and are inefficient in utilizing the available network and PCIe interconnect and memory subsystems, showing only up to 40\% peak efficient checkpointing throughput across data-parallel replicas. Approaches such as DeepFreeze~\cite{nicolaeDeepFreezeScalableAsynchronous2020}, TorchSnapshot~\cite{TorchSnapshot}, and LightCheck~\cite{lightCheck} attempt to mitigate the checkpointing overheads by both overlapping transfers with training and partitioning checkpoints across data-parallel replicas, but do not support hybrid pipeline, tensor, and data-parallel training setups.

\paragraph*{\textbf{Checkpointing for LLMs}}
Several recent efforts specifically target checkpointing for LLMs and
focus on efficient asynchronous 2-phase CPU-based snapshotting and
lazy persistence.  However, the reported checkpointing throughputs are
far from saturating the network (50+~GB/s and PCIe (25+~GB/s)
links. For example, Gemini~\cite{wang2023gemini} reports 3.13~GB/s
checkpointing throughput (9.4~GB shard of GPT-100B takes about 3 seconds for checkpointing).  
REFT~\cite{wang2023reliableREFT} reports
38\% PCIe bandwidth utilization at 6~GB/s, while TRANSOM's
checkpointing engine~(TCE)~\cite{wu2023transom} reports achieving a
throughput of $\sim$1.2~GB/s. Microsoft's Nebula~\cite{ziqiwangOptimizeCheckpointPerformance2023} (closed source) reports achieving 1-4~GB/s (GPT2-XL checkpoint of 20.6~GB takes 5 seconds to checkpoint). FastPersist~\cite{wang2024fastpersist} decouples training and checkpointing with Python multiprocessing, but inherits substantial (de)serialization overheads via process handoff and primarily evaluates on node-local NVMe rather than stable shared storage.

\paragraph*{\textbf{High-Performance Data Management Runtimes}}
In HPC, transparent checkpoint/restart runtimes (e.g., BLCR~\cite{hargrove2006berkeley}, CheCUDA~\cite{takizawa2009checuda}) capture whole-process state, while application-directed systems (e.g., VELOC~\cite{nicolaeVeloCHighPerformance2019, VELOCGPU-HiPC22, GPUPrefetch-HPDC23}, FTI~\cite{parasyris2020checkpoint-fti-gpu}) and I/O engines (e.g., ADIOS~\cite{godoy2020adios}) provide multi-level buffering and asynchronous data movement. However, these runtimes are not tailored to LLM training's \emph{3D checkpoint heterogeneity} (\S~\ref{sec:motivation:3d-hetereogenity}), nor do they exploit the long immutable forward/backward phases to hide flushes while minimizing serialization and metadata overheads.

\begin{figure*}[t]
\centering
\minipage{0.32\textwidth}
    \centering
    \includegraphics[width=\linewidth]{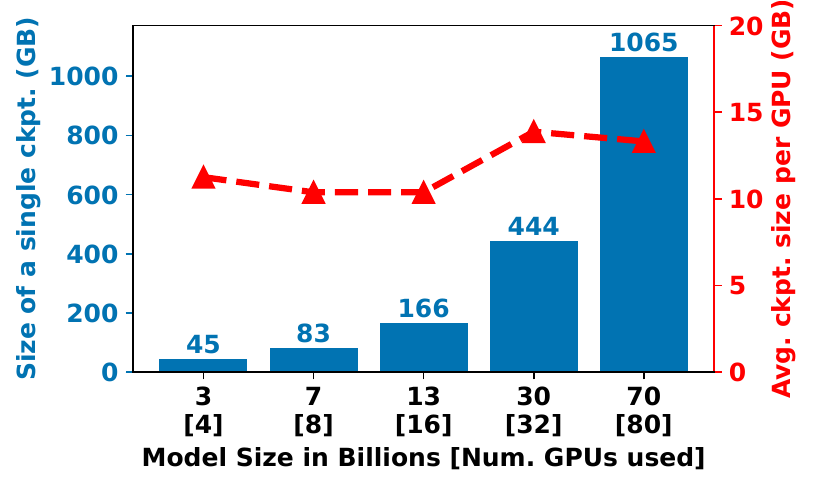}
    \caption{Aggregate checkpoint sizes of different model sizes and average checkpoint size per GPU.}
    \label{fig:agg-ckpt-size}
\endminipage\hfill
\hspace{2pt}
\minipage{0.32\textwidth}
    \centering
    \includegraphics[width=\linewidth]{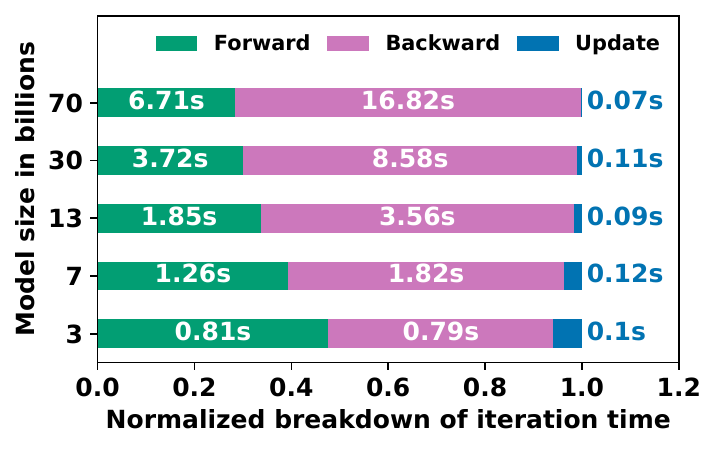}
    \caption{Different iteration phases. Model and optimizer states are immutable during forward and backward passes.}
    \label{fig:diff-phases-time}
\endminipage\hfill
\hspace{2pt}
\minipage{0.32\textwidth}
    \centering
    \includegraphics[width=\linewidth]{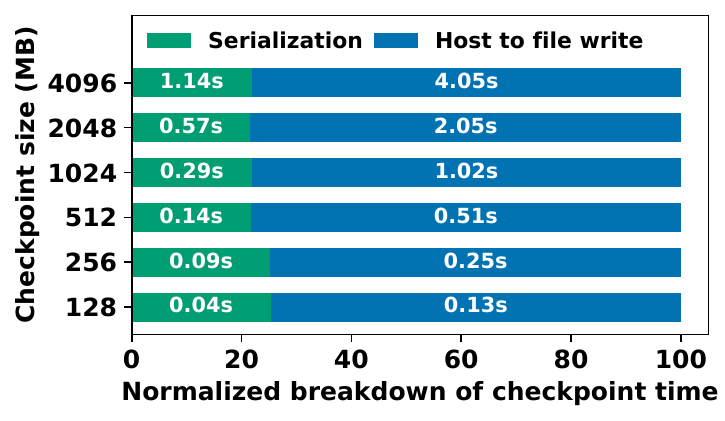}
    \caption{Breakdown of serialization and write performance for different data sizes.}
    \label{fig:serialize-write-perf}
\endminipage\hfill
\vspace{-10pt}
\end{figure*}

\section{Analysis of LLM Checkpointing Behavior}
\label{sec:motivation}

LLM checkpointing differs qualitatively from conventional DNNs along three axes: (i) \emph{large data volumes} (optimizer-dominated footprints), (ii) \emph{iteration structure} (immutable phases enabling safe overlap), and (iii) \emph{checkpoint composition} (many heterogeneous objects mapped to many files). We quantify these effects to motivate our proposed design choices.

\subsection{\textbf{Checkpoint Size Scaling and Load Balance}} 
\label{sec:motivation:load-balance}

Unlike SGD~\cite{ruder2016overview}, modern LLM training predominantly uses adaptive optimizers such as Adam~\cite{kingma2014adam}, which maintain per-parameter auxiliary state (e.g., momentum and variance) in addition to gradients and parameters. In addition to the model parameters, this large optimizer state is also essential to successfully restart training and cannot be omitted. As a result, checkpoint size is dominated by optimizer state and grows rapidly with model width: while checkpoint size increases roughly linearly with the number of Transformer layers, it increases \emph{quadratically} with hidden dimension ($d$) due to the $O(d^2)$ parameterization of attention/MLP matrices~\cite{rajbhandari2021zero}. Figure~\ref{fig:agg-ckpt-size} reports end-to-end checkpoint sizes for the models in Table~\ref{tab:models} trained with DeepSpeed (\S~\ref{sec:setup}); the near-constant checkpoint size per GPU across model scales indicates that the runtime shards states with good load balance (minor y-axis variation), but does not reduce the \emph{per-rank} volume that must be staged, serialized, and persisted.

\subsection{\textbf{Iteration Immutability Enables Safe Overlap}}
\label{sec:motivation:immutability}

To understand when checkpoint data can be safely extracted, we decompose each iteration into forward, backward, and optimizer updates. Figure~\ref{fig:diff-phases-time} shows that forward/backward dominate iteration time across model sizes; the update phase, which runs embarrassingly parallel computations for optimizers, e.g., ADAM, becomes comparatively small as communication and synchronization costs (pipeline/tensor send/recv and DP all-reduce) amplify with scale. Crucially, both model parameters and optimizer state are \emph{immutable} during forward and backward, and are only mutated during the update step. Therefore, GPU$\rightarrow$host staging can be issued asynchronously throughout forward/backward without coherence issues. Moreover, these DMA transfers primarily consume PCIe bandwidth, whereas training communication uses distinct fabrics (e.g., NVLink intra-node and GPUDirect RDMA inter-node), reducing contention between checkpoint staging and the critical-path communication of 3D parallel training.

\subsection{\textbf{3D Heterogeneity of Checkpoint Data Structures}}
\label{sec:motivation:3d-hetereogenity}

LLM training optimizes throughput by \emph{distributing and re-formatting application state} to match the GPU memory hierarchy and the parallel execution plan. These are not arbitrary checkpoint-engine choices: they are computational necessities that directly affect training efficiency, and any change in the underlying \emph{sharding} (data distribution) alters communication/computation balance and can degrade performance. In modern mixed-precision training~\cite{micikevicius2018mixedprecisiontraining, workshopBLOOM176BParameterOpenAccess2023}, forward/backward use FP16/BF16 activations/weights for speed and memory footprint, while numerically sensitive state (e.g., optimizer moments/variance and often master weights) is maintained in FP32. In addition, large-scale runtimes shard and optimize the layout of tensors also due to other aspects,
such as to reduce host/GPU memory allocation overheads and to enable efficient inter-GPU collective communication, which is needed because the parameters/gradients/optimizer partitions are split across TP/PP/DP ranks (and, with ZeRO, across DP replicas)~\cite{rajbhandari2021zero}. Beyond tensors, training also requires host-resident control state (Python dictionaries, namespaces, RNG seeds, configuration, and runtime metadata) to ensure correct restart and reproducibility. To reduce I/O overheads due to false sharing
under concurrency, many checkpointing runtimes collect and store independent shards and data types into separate files. 

For example, Table~\ref{tab:tensors-vs-non-tensors} shows the default distribution of the number of files, distribution of tensor vs non-tensor data structures, and varying precisions of tensors for a DeepSpeed training. We can observe directly \emph{3D checkpoint heterogeneity}: (1) \emph{residency}: GPU-resident tensors plus host-resident control/metadata, (2) \emph{type/precision}: FP16/BF16 and FP32 tensor payloads plus non-tensor objects, and (3) \emph{sharding/cardinality}: many independently owned tensor shards per rank whose boundaries are dictated by TP/PP/DP and optimizer sharding, not by checkpoint layout. While checkpoint files can be aggregated or reorganized, the shard boundaries primarily reflect the computational distribution; thus, checkpointing must efficiently capture many heterogeneous shards without perturbing the training layout. Existing state-of-the-art checkpointing engines~\cite{wang2024fastpersist, LightningAsyncCheckpointIO, TorchSnapshot} are oblivious to the \emph{3D checkpoint heterogeneity} of LLMs, lack capabilities for parallel streaming of large contiguous data objects (e.g., tensors), and perform data-oblivious serialization, which we investigate next.

\vspace{-0.5em}
\begin{table}[t]
    \centering
    \caption{3D checkpoint heterogeneity for LLama~2 LLMs with variable parameter size (3B, 7B,13B) and DP=1.}
    \setlength{\tabcolsep}{3.5pt}
    \begin{tabular}{|l||l c c c|}
        \hline
        \multirow{2}{*}{\textbf{Models}} & \multirow{2}{*}{\textbf{Files/Type}} & \textbf{Metadata} & \textbf{Parameters} & \textbf{Optimizer}\\
        & & \textbf{(Host)} & \textbf{(GPU)} & \textbf{(GPU)}\\
        \hline \hline
        \multirow{3}{*}{\shortstack{3B\\(TP=4,PP=1)}}  & \# of files      & 4   & 132 & 4   \\
                                                     & tensors      & 20~KB& 5.8~GB (FP16) & 35~GB (FP32) \\
                                                     & non-tensors  & 20~MB& 28~KB & 102~KB \\
        \hline
        \multirow{3}{*}{\shortstack{7B\\(TP=4,PP=2)}}  & \# of files      & 8   & 140 & 8   \\
                                                     & tensors      & 40~KB& 13~GB (FP16) & 82~GB (FP32) \\
                                                     & non-tensors  & 40~MB& 32~KB & 116~KB \\
        \hline
        \multirow{3}{*}{\shortstack{13B\\(TP=4,PP=4)}} & \# of files      & 16  & 172 & 16  \\
                                                     & tensors      & 80~KB& 25~GB (FP16) & 148~GB (FP32) \\
                                                     & non-tensors  & 80~MB& 40~KB & 160~KB \\
        \hline
    \end{tabular}
    \label{tab:tensors-vs-non-tensors}
\end{table}

\subsection{\textbf{Type-Agnostic Serialization and Coarse Transfers}}
\label{sec:motivation:serialization}

Checkpointing ultimately persists a \emph{logical} Python object (typically a nested \texttt{dict}) that mixes tensors and non-tensor metadata. However, serialization is not inherently required for a large fraction of this state: contiguous tensors and many primitive containers already expose byte-addressable buffers that can be written directly. In contrast, the dominant practice-- \texttt{torch.save} (also used by asynchronous engines, e.g., FastPersist~\cite{wang2024fastpersist})-- is type-agnostic: it traverses and serializes the entire object graph (using deep copies) even when most payload bytes are already 
contiguous, thus generating high yet unnecessary per-checkpoint serialization overheads.

To isolate this cost, we checkpoint a Python \texttt{dict} containing a host-resident contiguous tensor of varying sizes (excluding GPU$\rightarrow$host effects) and decompose time into serialization versus file write under \texttt{torch.save}. Figure~\ref{fig:serialize-write-perf} shows that serialization contributes a large and nearly size-invariant fraction of end-to-end checkpoint time ($\sim$22\%), indicating that redundant object-graph serialization is a first-order bottleneck and must be avoided to approach hardware limits.

Moreover, \texttt{torch.save} exhibits poor write-path efficiency: despite $\approx$10~GB/s node-level peak write capability on Polaris (\S~\ref{sec:setup}, Figure~\ref{fig:scale-microbench-data-sizes}), its effective flush throughput reaches only $\approx$1~GB/s, i.e., $\sim$10\% of peak (Figure~\ref{fig:serialize-write-perf}), reflecting coarse-grained transfers, buffering/copy overheads, and synchronous file semantics. TorchSnapshot~\cite{TorchSnapshot} increases write parallelism via chunked asynchronous I/O, but chunk-to-file mapping inflates file counts and exacerbates PFS metadata overheads~\cite{Gossman2026LLMCheckpointIO}. These results motivate a data-type aware pipeline that (i) bypasses serialization for byte-addressable tensor buffers and (ii) sustains high-throughput streaming writes without exploding metadata operations.

\section{\proj: System Design}
\label{sec:designandimpl}

Our goal is to design scalable multi-level asynchronous checkpointing solution that: (1) captures a globally consistent checkpoint of LLMs that includes all objects and data structures (distributed across GPU and host memory) of all workers corresponding to both the model parameters and the optimizer state (which are needed to successfully restart the training); (2) is aware of the \emph{3D heterogeneity} of checkpoint datastructures and performs selective serialization of only required objects in overlapped fashion with meta-data generation and I/O operations; (3) maximizes the checkpointing throughput to reduce the amount of time during which the training is blocked by checkpointing; and (4) minimize the resource contention and interference between the training and the asynchronous checkpoint process to reduce the end-to-end training duration.

\subsection{\textbf{Design Principles}}
\label{sec:design}

Based on the observations outlined in \S~\ref{sec:motivation}, we
introduce a series of high-level design principles that we adopt in
\proj to address the limitations of state-of-the-art LLM checkpointing
runtimes.

\begin{figure*}[t]
    \centering
    \includegraphics[width=0.9\linewidth]{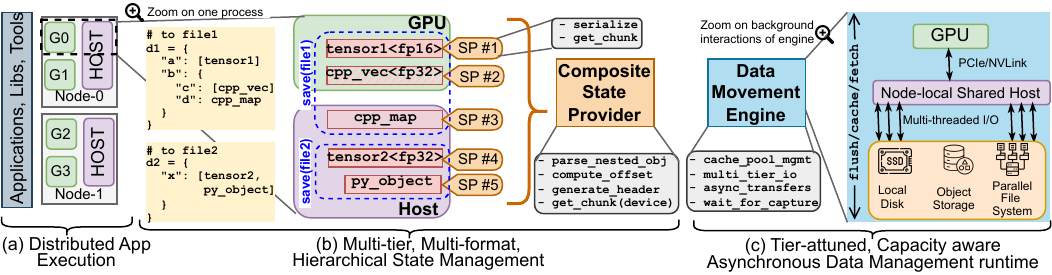}
    \caption{Overview of \proj: Composable state providers capture data structures
    subject to ``3D checkpoint heterogeneity'' in a streamlined fashion and flush them to multi-level storage tiers using a checkpoint engine.}
    \label{fig:high-level-design}
\end{figure*}

\subsubsection{\textbf{Coalescing of GPU Model/Optimizer Shards to Host Memory}}
\label{sec:design:principles:coalesce}

Conventional asynchronous multi-level checkpointing techniques (as
implemented in the related works mentioned in \S~\ref{sec:related})
move the checkpoints one at a time through the storage levels: first
they allocate host memory to hold the checkpoint, then they capture
the checkpoint on the host memory by performing a GPU-to-host copy,
then they asynchronously flush the checkpoint from the host memory to
persistent storage. If another checkpoint request arrives before the
previous checkpoint is finished flushing, it will be blocked waiting
for the flushes to complete. For small learning models that fit in the
memory of a single GPU, such an approach works reasonably well because
all model parameters and the optimizer state can be captured at once
in a single file. However, the combination of 3D parallelism and
optimizer state sharding targeted by our checkpointing scenario
results in many independent shards per GPU that correspond to both the
model parameters and the optimizer state. Eventually, each of these
shards need to be flushed to persistent storage, typically as a
separate file, as illustrated in Figure~\ref{fig:ckpt-sharding}(c)
and Table~\ref{tab:tensors-vs-non-tensors}.

In this case, conventional asynchronous multi-level approaches, e.g., CheckFreq~\cite{mohanCheckFreqFrequentFineGrained}, would
serialize the checkpointing of the shards. For example, if we
consider three shards in a checkpoint, two of which correspond to
layers $L1$ and $L2$ and the third corresponds to the optimizer state
shard, then only the flushing of the optimizer state shard will
overlap with the next iteration (forward pass, backward pass, and
updates), while the rest of the operations (allocate, copy, serialize, flush
$L1$; allocate, copy, serialize, flush $L2$; allocate, copy, serialize optimizer state) are synchronous. This severely degrades the performance of asynchronous
checkpointing to the point where it may become slower than synchronous
checkpointing.  

To optimize and extend the conventional asynchronous
multi-level checkpointing approach for multi-layered LLMs, approaches such as PyTorch's TorchSnapshot~\cite{TorchSnapshot}, illustrated in
Figure~\ref{fig:overlap-llm-flow}(b), can be used --- all the three
shards in the checkpoint ($L1$, $L2$, and optimizer) can be first
\emph{snapshotted} quickly using device-to-host copies.
Once the
snapshot of all layers involved in the checkpoint is complete, they
can be persisted through asynchronous flushes from the host to
disk. However, even such an advanced asynchronous approach slows down
training due to blocking GPU to host transfers, and inefficient serialization (as
evaluated in Table~\ref{tab:checkpoint-times-breakdown}).

To mitigate this issue, we propose three optimizations. First, we
pre-allocate enough host memory to hold all shards on the host
memory. This pre-allocated memory will be reused for all checkpoint
requests, effectively eliminating the allocation overheads for all
shards, both belonging to the same and different checkpoints. Second,
we pre-pin the allocated host memory, which accelerates GPU-to-host
data transfers, again for all shards of both the same and different
checkpoints. Third, we coalesce the copies of the different objects 
in each shard to host
memory, which eliminates the need to wait for the flushes of the
shards belonging to the same checkpoint to finish before initiating
more GPU-to-host copies.

\subsubsection{\textbf{Lazy Non-Blocking Capture Overlapping with Forward and Backward Pass}}
\label{sec:design:principles:lazy}

We leverage a key observation that the model and optimizer shards on
each GPU remains immutable during the forward pass and the backward
pass, and are updated later in bulk (typically through
\verb|optimizer.step()| for adaptive optimizers such as Adam~\cite{kingma2014adam}). Therefore,
unlike conventional asynchronous multi-level checkpointing techniques,
there is no need to block the next training iteration until a full
copy of the checkpoint is available on the GPU or host memory.  Instead, we
allow the next training iteration to start immediately after the
checkpoint request, and proceed to copy the shards to the host memory
while the forward pass and the backward pass are progressing in
parallel. Only when the update phase is about to begin, if the shard
copies on the host memory are not finished, then we delay the update phase
until they are finished. Furthermore, the flushes from the host memory
to persistent storage are also allowed to overlap with the update
phase. It is for this reason that we refer to our technique as
``lazy'' non-blocking copies: unlike copy-on-write, we start the copies
as soon as possible. At the same time, just like copy-on-write, if a 
modification is about to happen to the parameters (during the update
phase) and the copies have not finished, we block until the copies have
finished to guarantee consistency. An example is
illustrated in Figure~\ref{fig:overlap-llm-flow}(c,d): the forward and
backward pass of the second iteration $F2$ and $B2$ proceed
immediately after the first iteration has finished, at which point a
checkpoint request was issued. They overlap with the GPU-to-host
copies. The update phase $U2$ is delayed until the GPU-to-host copies
have finished to preserve consistency. This step can be omitted if
the copies have already finished. Meanwhile, the
previously captured checkpoints on the host are asynchronously flushed
to persistent storage. Finally, if the host memory that is reserved
for checkpointing is full, then the next checkpoint request needs to
wait for previous tensors to get evicted from the host memory after
they are flushed to the persistent storage, e.g., node-local NVMe
storage or parallel file system. We enforce this wait in order to
avoid running out of the host memory since GPU-to-host copies are
faster than host copies to persistent storage.

\subsubsection{\textbf{Composable Distributed State Providers}}

While the aforementioned design principles accelerate bulk I/O, checkpoint engines remain agnostic to the \emph{semantics} of the data they move, i.e., data type, layout, device residency, and the precise (de)serialization needs. To address the ``3D data heterogeneity'' discussed in \S~\ref{sec:motivation:3d-hetereogenity}, we introduce \emph{state providers (SPs)}: a lightweight abstraction that sits between the training runtime and the data movement engine (Figure~\ref{fig:high-level-design}). State providers present a uniform \emph{stream-oriented} view to the engine while isolating per–data-structure knowledge and policies about composition (e.g., coalescing fragmented chunks), (de)serialization, placement, and mapping to different files.
Thanks to this approach, the data movement engine can be agnostic to heterogeneity and
focus on a simple goal: to optimize multi-level I/O strategies based on competing
checkpoint data streamed as sequences of bytes by concurrent state providers.
For example, as depicted in Figure~\ref{fig:high-level-design}, each rank of
a distributed training (attached to a GPU) can specify different data structures to be saved into different files (potentially by different codes and modules). These
write operations are intercepted by state providers, each of which is specialized
to aggregate, reorganize, and serialize the data structures if needed as a stream
of bytes. Tensor can reside on either GPU or host memory and does not need to be
serialized. Dictionaries and Python objects need to be serialized in custom binary 
or JSON/pickle formats. State providers hide these details from the data movement engine
through a simple iterator that presents the next chunk in the stream (i.e., how many bytes from where the checkpoint engine can read). By composition, state providers
can be merged hierarchically to form a single stream that acts as a parallel producer
of checkpoint chunks in different memory tiers. The resulting composite state provider 
is primarily responsible for (a) computing object sizes and offsets in the stream
(which may not be known a priori for data structures that need to be serialized)
in order to generate meaningful chunks; (b) deciding how to group objects together and how to obtain a persistent format subject to a well-defined layout (e.g., what chunks need to be contiguous and in what file they need to be written), which is given as a hint to the data movement engine; (c) manage competition between state providers on different
memory tiers in a hierarchic 
fashion. Since the state providers simply return a memory view for pre-serialized objects, such as tensors, contiguous arrays, etc., they eliminate the unnecessary serialization overheads incurred by the state-of-the-art approaches (described in \S~\ref{sec:motivation:serialization}).

\subsubsection{\textbf{Streamlined Multi-level Flushing to Persistent Storage}}
\label{sec:design:principles:streamlined}

The data movement engine is responsible for implementing scalable 
I/O strategies for asynchronous multi-tier flushing of the checkpoint data
into a persistent format. The key to doing so efficiently is taking
advantage of the stream-oriented composition of the state providers,
who can expose chunks on GPU and host memory (either the original bytes or a serialized copy) in parallel. Since the chunks do not need to represent entire objects
(e.g., some tensors grow to huge sizes), the data movement engine can
start flushing an object from the original source to the rest of the storage
tiers as soon as it is partially available and some of its chunks have been
exposed, without having to wait for the full object to be processed by the state providers. Using this approach, separate physical paths (e.g., PCIe GPU-to-host, PCIe host-to-SSD, RDMA SSD-to-PFS) can be used in parallel to transfer the checkpointing data,  which reduces the I/O overheads associated with checkpointing. Furthermore, it is important
to note that GPUs have a separate GPU-to-host hardware copy
engine. Therefore, the memory accesses on a GPU issued during the
forward pass and the backward pass, regardless of whether to run
computational kernels or to communicate with other remote GPUs
(through NVLinks and/or GPUDirect RDMA),
do not compete with the copies of the chunks. Likewise, flushing from
host memory to persistent storage uses an entirely different I/O path
that does not interfere with the GPUs. As a consequence, our approach
maximizes the use of the I/O paths needed for checkpointing 
while minimizing interference due to competition. Thanks to this approach, 
except for unavoidable waits due to lazy non-blocking capture, training
iterations can effectively progress almost undisturbed by
checkpointing.

\subsubsection{\textbf{Overlapping I/O with Serialization}}
\label{sec:design:principles:metadata}

The separation between state providers and the data movement engine as a stream-oriented producer-consumer pattern has another important advantage. State providers have the
flexibility to decide how to serialize the checkpoint objects at what granularity,
which can produce many small chunks or fewer larger chunks. For objects that are
large and do not require serialization, large chunks can be exposed to the data movement
engine at a high rate immediately. Conversely, for objects that require serialization,
chunks can be produced later and the size can be adjusted to be smaller, in order to
increase the rate at which they are exposed. Since LLM checkpoints always include
a large number of huge tensors, by starting with the chunks of these tensors, the serialization and/or metadata overheads incurred by the state providers overlap 
with the I/O operations performed by the data movement engine. Furthermore, there
are enough large tensors to keep the data movement engine busy for a long time,
which is more than sufficient to allow the serialization of the other, relatively smaller data structures (typically holding metadata and small state information) to finish.
Thus, the data movement engine never has to wait for the state providers to expose
chunks. Using this approach, we effectively increase the overall checkpoint throughout compared with state-of-the-art approaches that do the opposite: they serialize the metadata and non-tensor objects first in a blocking fashion (to precompute an optimized persistent checkpoint layout from the beginning), and then proceed to flush the tensors. It is important to note, though, just like state-of-the-art approaches, we also need to 
store the checkpoints in an optimized persistent layout. This is non-trivial given
the dynamic nature of chunks. To this end, we adopt an hybrid fixed-offset, concurrent-log-structured append strategy: (1) we know how large the tensors are, therefore we can precompute offsets at which they need to be persisted in files; (2) we don't know how large serialized objects are, but we can interleave their chunks using a concurrent log-structured append pattern starting with the offset at which the tensors end; (3) we can finally append a metadata header at the end of the file to describe the layout and allow recovery of both serialized and non-serialized objects. Thus, we effectively overlap
I/O and serialization while optimizing the persistent checkpoint layout.

\subsection{\textbf{\proj Integration with Training Runtime}}\label{sec:arch}

\begin{figure*}[t]
    \centering
    \includegraphics[width=0.9\linewidth]{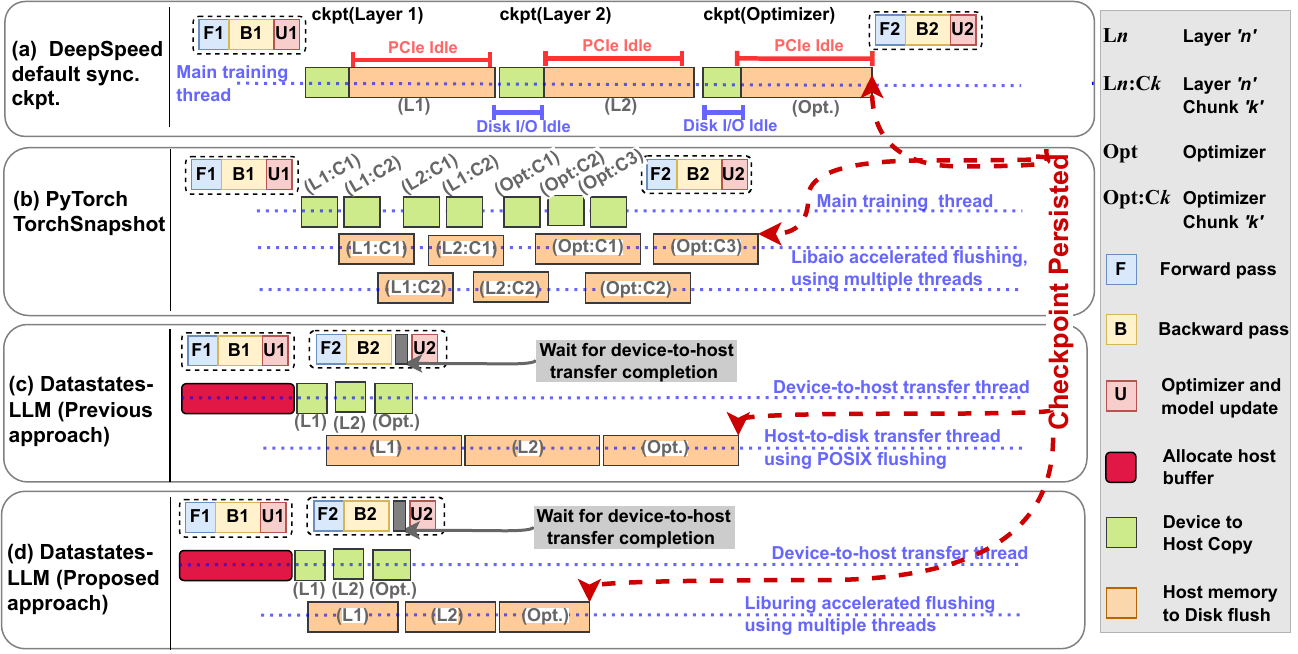}
    \caption{Overlapping LLM training with checkpointing using different approaches.}
    \label{fig:overlap-llm-flow}
\end{figure*}

We implement \proj as a drop-in checkpoint-engine backend for DeepSpeed, replacing the default synchronous \verb|torch.save()|-based engine. \proj is enabled via the DeepSpeed JSON configuration and requires a single tunable parameter: the per-process host-side pinned buffer capacity used as an intermediate checkpoint cache. The engine preserves DeepSpeed's asynchronous checkpoint API/semantics (based on FastPersist~\cite{wang2024fastpersist}). No model or application code changes are required as \proj is integrated in the DeepSpeed runtime~\footnote{\url{https://www.deepspeed.ai/tutorials/datastates-async-checkpointing/}}.

\subsection{\textbf{Implementation}}
\label{sec:impl}
\proj is implemented in C++/CUDA and exposed to DeepSpeed via thin Python bindings.\footnote{Source code: \url{https://github.com/DataStates/datastates-llm}.} The host cache is a lightweight pinned-memory circular buffer that enforces the producer-consumer flushing pattern in \S~\ref{sec:design} and blocks background flushing or training (if needed) when the cache is saturated. GPU$\rightarrow$host staging is issued on dedicated CUDA streams, while host$\rightarrow$storage persistence runs on background worker threads to decouple PCIe DMA, serialization/packing, and file flush. For storage I/O, \proj uses \texttt{liburing} with \texttt{O\_DIRECT} to reduce syscall overhead and page-cache interference for large sequential writes. Implementing staging and persistence in C++ avoids the overheads of Python-thread/process-based checkpointing (e.g., CheckFreq~\cite{mohanCheckFreqFrequentFineGrained}, LightCheck~\cite{lightCheck}, FastPersist~\cite{wang2024fastpersist}), which often pass checkpoint objects through multiprocessing queues and trigger data-oblivious (de)serialization and extra copies, exacerbating the serialization costs quantified in Figure~\ref{fig:serialize-write-perf}. Finally, \proj's \emph{state-provider} abstraction and data-movement engine are modular and can be integrated with other checkpoint runtimes (e.g., VeloC~\cite{nicolaeVeloCHighPerformance2019}) while preserving the same asynchronous staging/persistence pipeline.

\section{Performance Evaluation}
\label{sec:evaluation}

\subsection{\textbf{Experimental Setup}}
\label{sec:setup}

\subsubsection{\textbf{Platform}}

We evaluate on ALCF Polaris~\cite{polaris_alcf}, a 560-node GPU testbed. Each node has a 32-core AMD Milan CPU, 512~GB DDR4 (4 NUMA domains), two 1.6~TB local SSDs, and four NVIDIA A100-40GB GPUs (160~GB total HBM). GPUs are interconnected via NVLink and attached to host memory over PCIe Gen4; peak per-GPU unidirectional bandwidth is 85~GB/s (D2D NVLink) and 25~GB/s (pinned D2H/H2D PCIe). Polaris has GPU--NUMA affinity (1:1), reducing PCIe/NUMA contention for concurrent D2H transfers. Persistent storage is a Lustre PFS~\cite{schwan2003lustre} with 160 OSTs and 40 MDTs (650~GB/s aggregate peak).

\subsubsection{\textbf{Software}}
Nodes run Cray SLES~15 with CUDA driver 565.57, NVCC 12.6.68, Python 3.12.11, PyTorch 2.5.1, and DeepSpeed 0.16.6. We use up to 64 nodes (256 GPUs) to study (i) scaling under TP/PP/DP, and (ii) PFS contention under concurrent checkpoint flushes.

\subsection{\textbf{Compared Approaches}}

\subsubsection{\textbf{DeepSpeed Default}}
DeepSpeed's default checkpointing uses PyTorch \verb|torch.save()|~\cite{rasleyDeepSpeedSystemOptimizations2020}. It blocks training and synchronously serializes/writes checkpoint shards to the PFS, providing a fully persistent checkpoint at each save point (illustrated as \emph{(a) DeepSpeed Default} in Figure~\ref{fig:overlap-llm-flow}.

\subsubsection{\textbf{TorchSnapshot}}
TorchSnapshot~\cite{TorchSnapshot} is a state-of-the-art PyTorch checkpoint runtime (Figure~\ref{fig:overlap-llm-flow}(b)). It works by (i) chunking tensors to pipeline device$\rightarrow$host and host$\rightarrow$disk transfers, and (ii) uses multi-threaded writes of chunk files to increase write-level parallelism. 

\subsubsection{\textbf{\proj-Old}}
Our prior DataStates-LLM engine~\cite{maurya2024datastates} overlaps checkpointing with forward/backward immutability and implements the coalescing, lazy persistence, and streamlined staging principles (\S~\ref{sec:design:principles:coalesce}, \S~\ref{sec:design:principles:lazy}, \S~\ref{sec:design:principles:streamlined}). This approach is an advanced representative of runtimes such as CheckFreq~\cite{mohanCheckFreqFrequentFineGrained} and FastPersist~\cite{wang2024fastpersist} due to asynchronous C++ based I/O (overcoming Python GIL limitations) and zero training-to-checkpoint cross-process serialization (\S~\ref{sec:motivation:serialization}).

\subsubsection{\textbf{\proj}}
\proj is the improved engine over \proj-Old~\cite{maurya2024datastates}, incorporating the full set of proposed design principles and optimizations (Figure~\ref{fig:overlap-llm-flow}(d)).

\subsection{\textbf{Evaluation Methodology}}
\label{sec:eval:method}
\begin{table}
    \centering
    \caption{Configuration of models and runtime used for evaluations
      derived from
      BLOOM 3B~\cite{workshopBLOOM176BParameterOpenAccess2023}
      and
      Llama~\cite{touvronLlamaOpenFoundation2023}.}
    \label{tab:models}
    \setlength{\tabcolsep}{5pt}
    \begin{tabular}{|c||>{\columncolor[gray]{0.8}}c|c|c|c|c|}
    \hline
    \textbf{Model Size in Billions} & \textbf{3} & \textbf{7} & \textbf{13} & \textbf{33} & \textbf{70} \\
    \hline
    \hline
    Layers & 30 & 32 & 40 & 60 & 80 \\
    Hidden dim. & 2560 & 4096 & 5120 & 6656 & 8192 \\
    Atten. heads & 32 & 32 & 40 & 52 & 64 \\
    Num. of nodes & 1 & 2 & 4 & 8 & 20  \\
    Tensor parallelism & \multicolumn{5}{|c|}{4 (=Number of GPUs per node)} \\
    Pipeline parallelism & \multicolumn{5}{|c|}{=Number of nodes} \\
    ZeRO optimization & \multicolumn{5}{|c|}{Stage 1 (Partition optimizer state)} \\
    \hline
    \end{tabular}
\end{table}

\subsubsection{\textbf{Models, Sharding, and Dataset}}

We evaluate five production-representative LLM configurations spanning BLOOM-3B~\cite{workshopBLOOM176BParameterOpenAccess2023}, Llama~33B, and Llama~7B, 13B, and 70B~\cite{touvronLlamaOpenFoundation2023} (Table~\ref{tab:models}). To match the typical node-level communication structure on Polaris, we fix tensor parallelism to TP$=4$ (one node) to exploit NVLink for intra-layer collectives. To fit models across distributed GPU memory, we split the model with pipeline parallelism (PP) across the number of nodes in Table~\ref{tab:models}, using DeepSpeed/Megatron's default uniform partitioning that balances trainable parameters per stage. Unless stated otherwise, DP$=1$ (single replica) to isolate checkpointing costs from replica scaling. For DP experiments, we enable DeepSpeed ZeRO-1 to shard optimizer state across replicas (Figure~\ref{fig:ckpt-sharding}(d)), which changes both per-rank checkpoint size and I/O parallelism.

We train on the OSCAR-en subset distributed with the BLOOM repository (79K records)~\cite{workshopBLOOM176BParameterOpenAccess2023}, tokenized with the Llama~2 tokenizer~\cite{touvronLlamaOpenFoundation2023}. We use a sequence length of 2048 and a micro-batch size of 16 across all configurations to avoid out-of-memory (OOM) errors in any configuration.

\subsubsection{\textbf{Memory and Storage Tiers}}
All approaches use a bounded pinned host cache of 80~GB per node; remaining host memory is reserved for dataloader/prefetch buffers and runtime/driver allocations. Given the observed per-GPU checkpoint shard sizes (10--15~GB in Figure~\ref{fig:agg-ckpt-size}), 80~GB/node is sufficient to hold (approximately) one full checkpoint version across the four GPUs per node, enabling overlap via host-side buffering without unbounded memory growth. Checkpoint shards are flushed from host memory directly to the Lustre PFS~\cite{schwan2003lustre}, which serves as stable shared storage.

\subsubsection{\textbf{Key Performance Metrics}}
Throughout our evaluations, we measure the following metrics for comparing the aforementioned approaches: (1) checkpointing throughput
of different model sizes to evaluate the blocking checkpointing
overhead on the application for increasingly complex
LLMs; (2) impact on iteration duration during checkpointing to evaluate the slowdown and interference caused by checkpointing on
training iterations; and (3) end-to-end training runtime to study the
broader impact on overall job completion times. We evaluate the above
metrics under different settings: (a) varying degrees of data
parallelism: this setting studies the
impact of strong scaling (more flushing bandwidth available to capture
the checkpoint of the same size from replicas), and (b) varying checkpointing
frequency, similar to gradient accumulation from the perspective of checkpointing, to study how the training performs for different degrees of
I/O pressure arising from frequent or sparse checkpointing scenarios.

\subsection{\textbf{Performance Results}} 
\begin{figure*}[t]
\centering
\minipage{0.32\textwidth}
    \centering
    \includegraphics[width=\linewidth]{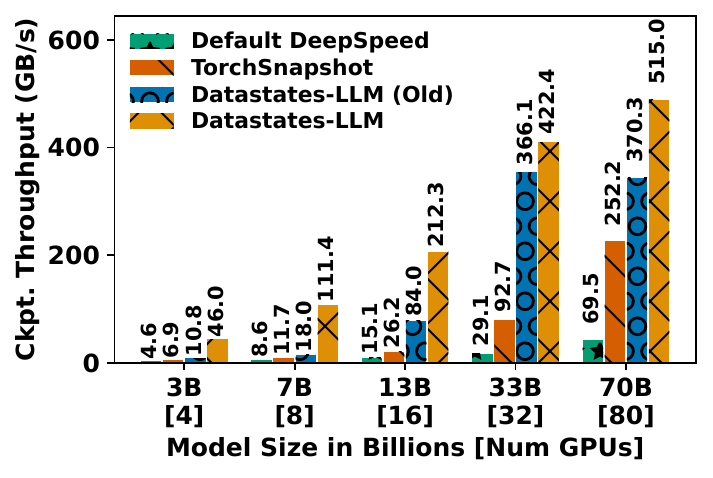}
    \caption{Aggregate checkpointing throughput for different model sizes. Higher is better.}
    \label{fig:ckpt-thru-diff-model-sizes}
\endminipage\hfill
\minipage{0.32\textwidth}
    \centering
    \includegraphics[width=\linewidth]{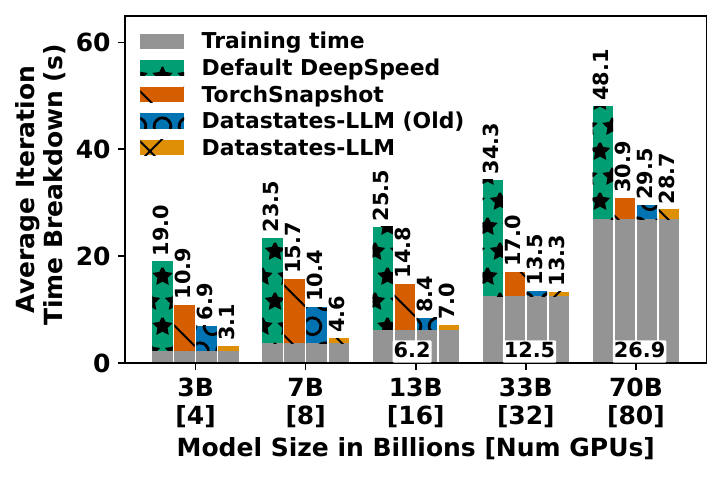}
    \caption{Average training iteration time for different model sizes when checkpointing. Lower is better.}
    \label{fig:per-process-iter-diff-model-sizes}
\endminipage\hfill
\minipage{0.32\textwidth}
    \centering
    \includegraphics[width=\linewidth]{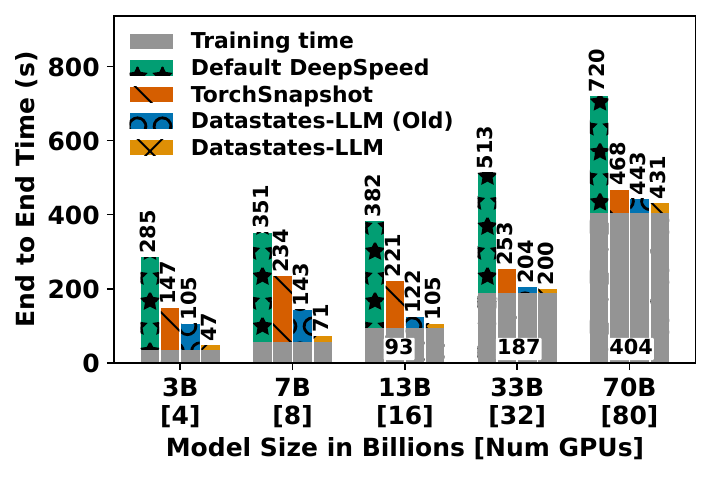}
    \caption{End-to-end training time for 15 iterations for different model sizes. Lower is better.}
    \label{fig:per-process-e2e-diff-model-sizes}
\endminipage\hfill
\end{figure*}

\subsubsection{\textbf{Scaling Model Sizes}}

We first evaluate checkpointing under increasing model size using two application-facing metrics. First, the \emph{effective checkpoint throughput}, defined as
\emph{global checkpoint size divided by the time for which the training is blocked by checkpointing}, i.e., time to initiate a checkpoint (including any synchronous serialization, header/metadata construction, and scheduling/staging setup) plus any additional blocking due to required synchronization before entering the optimizer update (e.g., waiting for the previous snapshot). Since checkpointing is a blocking collective after the update phase, the effective throughput is dictated by the slowest rank. Second, we evaluate the \emph{iteration duration under checkpointing}, which captures both \emph{direct} stall (snapshot capture/synchronization) and \emph{indirect} interference (forward/backward slowdown due to concurrent staging/persistence). We stress both effects by running for 15 iterations and checkpointing at every iteration.

Figure~\ref{fig:ckpt-thru-diff-model-sizes} shows that effective throughput increases with model size for all approaches because (1) larger models have longer iterations (Figure~\ref{fig:diff-phases-time}), providing slack to drain background flushes and reducing blocking on pending persistence; and (2) larger models span more nodes (Table~\ref{tab:models}), increasing aggregate PCIe staging bandwidth and PFS write concurrency. Despite similar scaling trends, \proj substantially raises the throughput envelope compared to DeepSpeed and TorchSnapshot; achieving at least $2\times$ and up to $10\times$ higher effective checkpoint throughput; and relative to \proj-Old, \proj improves by $1.2\times$--$7\times$.

Figure~\ref{fig:per-process-iter-diff-model-sizes} quantifies the training overhead by breaking per-rank iteration time into training (annotated at the bottom of bars) and checkpoint components. For smaller models (3B/7B/13B), checkpointing dominates iteration time under DeepSpeed and TorchSnapshot, indicating substantial blocking/interference. \proj reduces checkpoint time to negligible levels, achieving up to $2\times$ faster checkpointed iterations than \proj-Old, thanks to state provider-based serialization and kernel-accelerated flushing. For larger models, training (compute+communication) dominates iterations, but \proj still reduces checkpoint overhead, compounding to significant GPU-hour savings.

Finally, we measure end-to-end time for 15 iterations with per-iteration checkpoints to capture any backlog effects from accumulated asynchronous flushes. Figure~\ref{fig:per-process-e2e-diff-model-sizes} shows that \proj consistently reduces end-to-end time, matching the per-iteration trends in Figure~\ref{fig:per-process-iter-diff-model-sizes} and confirming that \proj sustains overlap without building an I/O tail.

\begin{figure*}[ht]
\minipage{0.32\textwidth}
    \centering
    \includegraphics[width=\linewidth]{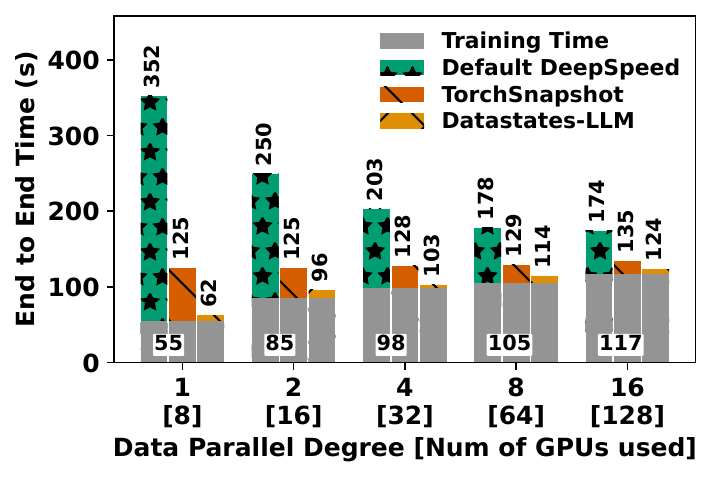}
    \caption{End-to-end training time for 15 iterations for the 7B model with increasing data parallelism. Lower is better.}
    \label{fig:scale-dp-7B}
\endminipage\hfill
\minipage{0.32\textwidth}
    \centering
    \includegraphics[width=\linewidth]{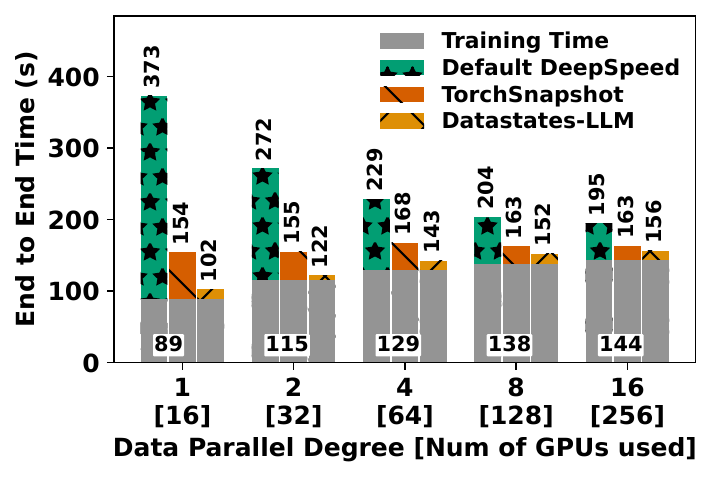}
    \caption{End-to-end training time for 15 iterations for 13B model with increasing data parallelism. Lower is better.}
    \label{fig:scale-dp-13B}
\endminipage\hfill
\minipage{0.32\textwidth}
    \centering
    \includegraphics[width=\linewidth]{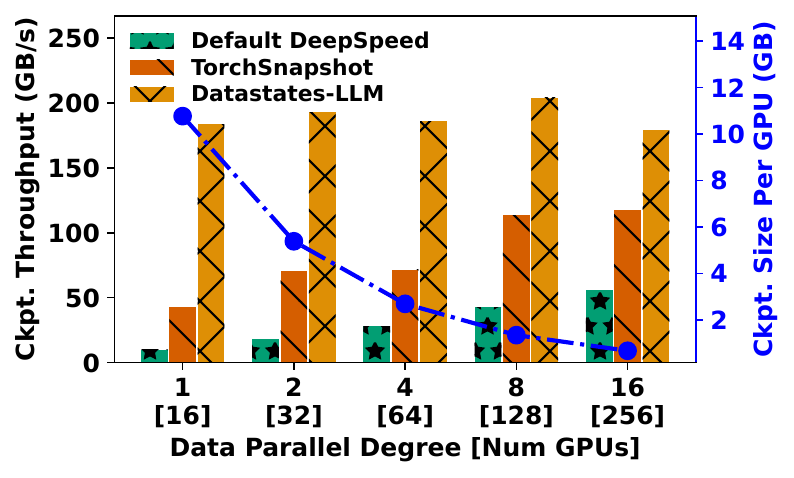}
    \vspace{0pt}
    \caption{Checkpointing throughput and size per GPU for the 13B model with increasing data parallelism.}
    \label{fig:scale-dp-13B-ckpt-time}
\endminipage\hfill
\end{figure*}

\subsubsection{\textbf{Fixed LLM Size with Increasing Data Parallelism}}
We next study checkpoint overheads under increasing data parallelism (DP), which determines both (i) the degree of optimizer-state sharding (ZeRO-1) and hence per-rank checkpoint size, and (ii) the number of concurrent writers contending on the shared Lustre PFS. We run 15 iterations and checkpoint every iteration, scaling DP from 1 to 16 for the 7B and 13B models. Larger models (33B/70B) exhibit similar trends to 13B but require $>$256 GPUs and have a larger training-dominated iteration fraction (shown in Figure~\ref{fig:per-process-e2e-diff-model-sizes}). We cap DP at 16 (256 GPUs) since large DP replication is uncommon due to high costs( e.g., BLOOM-175B used DP$=8$ on 384 GPUs~\cite{workshopBLOOM176BParameterOpenAccess2023}). To focus on state-of-the-art comparisons under expensive large-scale runs, we omit \proj-Old in the remaining experiments and report observations from DeepSpeed, TorchSnapshot, and \proj.

\begin{figure*}[ht]
\minipage{0.32\textwidth}
    \centering
    \includegraphics[width=\linewidth]{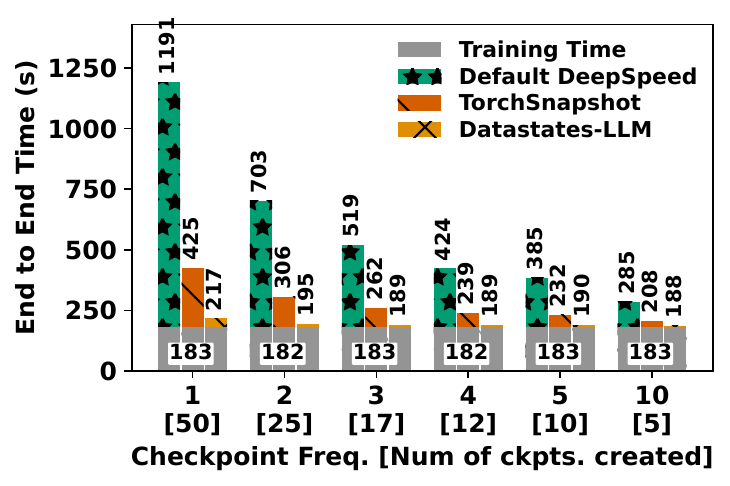}
    \caption{End-to-end training time for 50 iterations with different checkpointing intervals for 7B model. Lower is better.}
    \label{fig:scale-ckpt-interval-7B}
\endminipage\hfill
\minipage{0.32\textwidth}
    \centering
    \includegraphics[width=\linewidth]{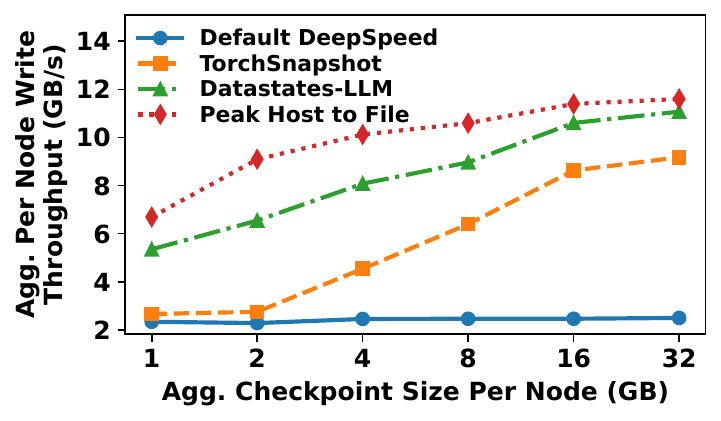}
    \vspace{0.6pt}
    \caption{Checkpointing throughput per node when scaling data sizes using different approaches. Higher is better.}
    \label{fig:scale-microbench-data-sizes}
\endminipage\hfill
\minipage{0.32\textwidth}
    \centering
    \includegraphics[width=\linewidth]{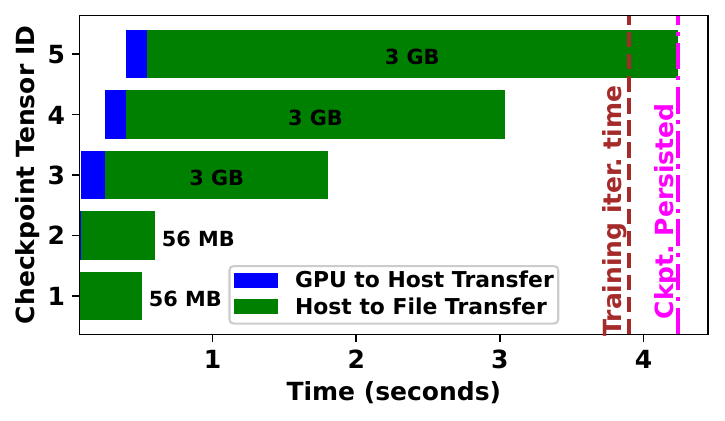}
    \vspace{0.6pt}
    \caption{Overlapping and streamlined checkpointing of selected tensors for 7B model with \proj on a GPU.}
    \label{fig:datastates-timeline}
\endminipage\hfill
\end{figure*}

Figure~\ref{fig:scale-dp-7B} and Figure~\ref{fig:scale-dp-13B} report end-to-end time versus DP for 7B and 13B. As DP increases, the annotated training component grows (near-linearly) due to added gradient averaging, yet \proj reduces end-to-end time by 1.3--5.7$\times$ across all DP scales, indicating that our checkpointing pipeline overlaps well with training, even as communication costs rise.

We also observe that the checkpointing time for all approaches reduces with increasing DP (shown by non-grayed bars) in Figure~\ref{fig:scale-dp-7B} and Figure~\ref{fig:scale-dp-13B} for 7B and 13B models, respectively. We zoom in on the 13B model in Figure~\ref{fig:scale-dp-13B-ckpt-time} and note two reasons for this effect: (1) smaller size to checkpoint per rank-- from 10.4~GB at DP=1 to 650~MB at DP=16, due to partitioned optimizer states (\S~\ref{sec:background}), shown by the minor y-axis; and (2) more nodes flushing in parallel (strong scaling). However, smaller per-rank payloads amplify fixed per-checkpoint costs (header creation, serialization, and asynchronous launch) and increase metadata pressure due to more (smaller) files, which limits scaling for existing engines. \proj sustains near-uniform effective throughput across DP by minimizing fixed overheads and maintaining a steady staging/persistence pipeline, translating strong-scaling checkpoint efficiency into lower end-to-end training time.

\subsubsection{\textbf{Increasing Checkpointing Frequency}}
Next, we vary the checkpoint frequency, i.e., iterations elapsed between consecutive checkpoints, which is also analogous to increasing gradient accumulation, i.e., multiple forward and backward passes over microbatches before running a single update on aggregated gradients, commonly used in practice~\cite{workshopBLOOM176BParameterOpenAccess2023,touvronLlamaOpenFoundation2023}. Larger checkpoint intervals provide more slack to complete asynchronous flushes to persistent storage and free up the host buffer for capturing subsequent checkpoints.

We train the 7B model for 50 iterations, and checkpoint at varying frequencies. We select the 7B model due to its fast forward-backward phases, which offer less overlap opportunity, making it a stress case for asynchronous checkpointing. Figure~\ref{fig:scale-ckpt-interval-7B} shows end-to-end time decreases with larger checkpoint intervals (fewer checkpoints, lower I/O pressure). Despite this, \proj maintains substantially lower runtime at all frequencies: it can take 5$\times$ more frequent checkpoints for comparable overheads to the best baseline-- \proj completes 50 iterations in 195 seconds when checkpointing every 2 iterations, while TorchSnapshot takes 208 seconds when checkpointing after every 10 iterations. Overall, \proj improves end-to-end runtime by 1.11$\times$-- 5.5$\times$ across checkpoint frequencies.

\subsubsection{\textbf{Ablation Studies}}
We next dissect checkpointing into \emph{blocking} (critical training path) and \emph{background} tasks. Blocking work constructs the capture plan and includes metadata/header construction, serialization and launching of asynchronous flush operations. Background work performs multi-tier flushing to persistent storage. We attribute overheads by isolating three sub-operations: (i) metadata computation \& serialization, (ii) GPU$\rightarrow$host staging, and (iii) host$\rightarrow$file persistence. We analyze a single checkpoint on one rank for the 7B model, which produces multiple files for distinct objects of the model and optimizer states (shown in Table~\ref{tab:tensors-vs-non-tensors}).

\begin{table}[t]
\centering
\small
\setlength{\tabcolsep}{2pt} %
\caption{Breakdown of different sub-operations during checkpointing on one rank for 7B model. Times in blue color represent background operations overlapped with training.}
\begin{tabular}{|l||ccc|}
\hline
\textbf{Sub-Operations} & \textbf{DeepSpeed} & \textbf{TorchSnapshot} & \textbf{Datastates-LLM} \\
\hline \hline
Metadata/Serialize & 3.9~sec. & 0.0258~sec. & 0.0156~sec. \\
GPU $\rightarrow$ Host      & 1.9~sec. & 1.3~sec. & \textcolor{blue}{0.6~sec.} \\
Host $\rightarrow$ File     & 16.1~sec. & \textcolor{blue}{11.5~sec.} & \textcolor{blue}{3.8~sec.} \\
\hline
\end{tabular}
\label{tab:checkpoint-times-breakdown}
\end{table}

\paragraph*{\textbf{Breakdown of Sub-Operations During Checkpointing}}
Table~\ref{tab:checkpoint-times-breakdown} aggregates per-checkpoint time across all files on one rank. DeepSpeed's \texttt{torch.save} serializes the full Python object graph (including pre-serialized tensor payloads), yielding high metadata/serialization cost. TorchSnapshot and \proj parse the checkpoint object to directly persist tensor-like buffers (tensors, numpy arrays, etc.) and serialize only the residual object; \proj further reduces blocking overhead via state-provider-based lazy header construction and streamlined serialization (\S~\ref{sec:design:principles:metadata}), improving metadata/serialization by 1.65$\times$ over TorchSnapshot. For GPU$\rightarrow$host, DeepSpeed and TorchSnapshot stage conservatively with blocking copies and non-pinned buffering, whereas \proj uses asynchronous DMA into a reusable pinned host cache (memory-pool backed), reducing staging time (overlapped with training). For host$\rightarrow$file, DeepSpeed relies on single-threaded \texttt{torch.save} flushing; TorchSnapshot and \proj use multi-threaded persistence, but \proj is still $3\times$ faster than TorchSnapshot on this phase, which we attribute to the flushing strategy investigated next.

\paragraph*{\textbf{Microbenchmarking I/O Performance}}
To isolate host$\rightarrow$file throughput, we microbenchmark concurrent flushing from 4 ranks (4 GPUs/node) on a single node for increasing tensor sizes. Each engine checkpoints a single GPU-resident tensor (capturing GPU$\rightarrow$host plus host$\rightarrow$file behavior); we also report an ``ideal'' host-only baseline that omits GPU$\rightarrow$host transfers and represents peak flush capability. Figure~\ref{fig:scale-microbench-data-sizes} shows aggregated node-level throughput rises with size and saturates beyond $\sim$4~GB/GPU (16~GB/node). DeepSpeed remains low due to non-optimized single-threaded flushes. TorchSnapshot and \proj improve with size via concurrent flushing but fall behind the host-only peak due to GPU$\rightarrow$host staging overheads; across all sizes, \proj outperforms TorchSnapshot by 1.25$\times$--2.5$\times$.

\paragraph*{\textbf{Streamlining of Multi-Tier Flushing}}
To understand the end-to-end flushing behavior of \proj, we study the overlapping and streamlined transfer of tensors that compose the major volume of checkpoints across multiple memory and storage tiers. We consider the 7B model as a representative model, and zoom in on a single rank. The checkpoint on a single rank produces 20 files, composed of 106 tensors with sizes ranging from 8~KB to 3.5~GB. Figure~\ref{fig:datastates-timeline} shows the timeline for the 5 largest of the 106 tensors. We observe that \proj streamlines transfers from GPU to host, such that each object can use peak PCIe throughput, and then multiple threads from the host overlap the flushes of these tensors to different files in parallel. Such asynchronous multi-level streamlining and overlapping of transfers significantly reduces the perceived I/O overheads by the training runtime.

\section{Conclusions}
In this work, we target the dominant checkpointing overheads in large-scale distributed LLM training under hybrid (TP/PP/DP) parallelism in widely used runtimes such as DeepSpeed. We show that existing LLM-oriented checkpoint engines still incur significant training stalls because they (i) fail to systematically exploit iteration semantics, specifically the immutability of model/optimizer state during forward/backward, to safely overlap staging and persistence, and (ii) leave substantial PCIe, memory, and storage bandwidth underutilized due to synchronous and data-oblivious host-staging, serialization, and coarse-grained flushing. We design and implement \proj, a transparent checkpoint engine that overlaps checkpoint I/O with immutable phases by combining: coalesced GPU$\rightarrow$host staging of sharded states; lazy non-blocking snapshot capture; streaming multi-level persistence to the PFS; hierarchic, distributed, and heterogeneous state providers; and overlapping I/O and metadata generation with the required serialization operations. Across production-representative BLOOM/Llama configurations, DP scaling, and checkpoint frequencies, \proj improves checkpoint throughput by 3$\times$--4.2$\times$ over state-of-the-art baselines and reduces end-to-end training time by 1.3$\times$--2.2$\times$.

Future work will extend \proj with data-reduction (differential checkpointing, compression) to further lower network/storage costs at high checkpoint rates, support offloaded model/optimizer states across deeper memory tiers, and investigate shard aggregation/consolidation to mitigate PFS metadata pressure without sacrificing parallelism.

\section*{Acknowledgements}
This work is supported in part by the U.S. Department of Energy (DOE),
Office of Science, Office of Advanced Scientific Computing Research
under contract DEAC02-06CH11357/0F-60169 and the National Science
Foundation (NSF) under award no.\ 2411386/2411387, 2106634/2106635, 2514056. Results
in this paper are obtained using ALCF HPC systems, and NSF
Cloudlab and Chameleon testbeds.

\bibliographystyle{IEEEtran}
\bibliography{references}
\end{document}